\documentclass[prl, twocolumn, showpacs]{revtex4-1}
\usepackage{amssymb}
\usepackage{amsmath}
\usepackage{epsfig}
\usepackage{bm}
\usepackage{xcolor}
\usepackage[unicode=true,colorlinks=true]{hyperref}

\begin{document}

\title{Effective one-band approach for the spin splittings in quantum wells}

\author{P.~S.~Alekseev}
\author{M.~O.~Nestoklon}
\affiliation{Ioffe Institute, St.~Petersburg 194021, Russia}

\begin{abstract}

The spin-orbit interaction of 2D electrons in the  quantum wells
grown from the III-V semiconductors consists of the two parts
with different symmetry: the Bychkov-Rashba and the Dresselhaus
terms. The last term is usually attributed to the bulk spin-orbit
Hamiltonian which reflects the $T_d$ symmetry of the zincblende
lattice. While it is known that the quantum well interfaces may also
contribute to the Dresselhaus term,  the exact structure and 
the relative importance of the interface and the bulk contributions
are not well understood yet. To compare the bulk
contribution with the interface one, we perform tight-binding
calculations of the spin splittings of the electron levels in [100]
GaAs/AlGaAs quantum wells and analyze the obtained spin splittings
within the one-band effective mass electron Hamiltonian containing
the two interface contributions to the Dresselhaus 
term. We show that the dependencies of the spin splittings on the
quantum well width and the electric field along the growth direction
are perfectly reproduced by the analytical one-band
calculations and the magnitude of the interface contribution to the
spin-orbit interaction for sufficiently narrow quantum wells is of
the same order as the contribution from the bulk Dresselhaus
Hamiltonian.
\end{abstract}


\maketitle

\section{Introduction}
The spin-orbit interaction of two-dimensional  (2D) electrons in the
heterostructures based on the non-centrosymmetric cubic
semiconductors has been extensively investigated during the last
three decades \cite{Winkler_book,Zutic04}. However, there is still
no complete understanding of the physical nature and magnitudes of
different contribution to the spin-orbit interaction even in the
conventional GaAs/AlGaAs heterostructure systems.

For the bulk III-V semiconductors the one-band effective mass
electron Hamiltonian with the spin-orbit terms and the
effective-mass anisotropy can be derived from the 14-band Kane {\bf
k}$\cdot${\bf p}-Hamiltonian \cite{Pikus}. The magnitude of  the
spin-orbit term in the one-band electron Hamiltonian can be
obtained from the gaps between the different bands of a
semiconductor and the interband matrix elements of the momentum
operator  between the different bands in the {\bf k}$\cdot${\bf
p}-Hamiltonian.

The spin-orbit interaction of 2D electrons in the
heterostructures grown from the III-V semiconductors consists of the
two parts with different symmetry: Bychkov-Rashba and Dresselhaus
terms. The isotropic part of the spin-orbit interaction (the
Bychkov-Rashba contribution) is induced by electric field along the
growth direction. It consists of the two contributions: the bulk
contribution, associated with the smooth part of the electric field
along the normal, and the interface contribution, associated with
the strong atomic field at the well interfaces (i.e., the electric
fields due discontinuities of the band edges  at the interfaces)
\cite{Rashba_from_Alekseev}. Analogously, the anisotropic part
of the spin-orbit interaction of 2D electrons (the Dresselhaus term)
contains the bulk contribution, related with the spin interaction in
the zincblende lattice, and the interface contribution, determined
with the atomic structure of the interfaces.

For the first time, the interface anisotropic terms in the effective
mass Hamiltonian were proposed in Refs.~\cite{Ivchenko96,Krebs98}
for the case of the hole Hamiltonian in  [100] quantum wells in
order to describe the unusual optical properties of that
structures. In the papers \cite{Vervoort99,Rossler} it was shown
that the Hamiltonian of Refs.~\cite{Ivchenko96,Krebs98} leads to a
spin-orbit interface anisotropic term in the effective electron
Hamiltonian, which induces  a contribution to the electron spin
splittings, additional to the bulk contribution. More recently, the
two interface anisotropic spin-dependent terms were introduced in
the electron effective Hamiltonian in
Refs.~\cite{Alekseev1,Alekseev2} in an attempt to describe lateral
anisotropy of 2D electron $g$-factor 
recently observed in the
 [100] GaAs quantum wells \cite{Nefyodov10,Nefyodov11_1,Nefyodov11_2}.
Similar interface terms were derived in Ref.~\cite{13} from the {\bf
k}$\cdot${\bf p}-Hamiltonian containing infinite number of bands.
The analysis of the experimental  data from Refs.
\cite{Nefyodov10,Nefyodov11_1,Nefyodov11_2} shows  that the
contributions to the $g$-factor anisotropy from the quantum well
interfaces and from the bulk regions are of the same order of
magnitude \cite{Alekseev2}. Recently, it was demonstrated
within the framework of the 14-band Kane model that the interface
spin-orbit terms are substantial in the Luttinger $4 \times
4$ Hamiltonian  for 2D holes in GaAs quantum wells \cite{Glazov}.

It should be mentioned that, at present, there is a strong
controversy concerning the value of the bulk  spin-orbit constant
$\gamma$ in various semiconductors
\cite{Pikus,11,12,Chantis06,Chantis08}. For example, it was
concluded in Ref.~\cite{11} from measurements of the spin splittings
of the electron dispersion in GaAs quantum wells that the bulk
spin-orbit constant $\gamma$ in GaAs is approximately half  of the
value which was previously accepted in literature \cite{Pikus,12}.
However, in the interpretation of the
 experimental data in Ref.~\cite{11}, the presence of anisotropic spin-orbit terms
localized at the interfaces of the wells was not taken into account.
Some novel ways of determining the bulk spin-orbit
Dresselhaus parameter from the experiments on bulk semiconductors subjected in
homogenous and inhomogeneous magnetic field were discussed in
Refs.~\cite{Alekseev07,Alekseev08,Alekseev09,Alekseev15}.

The tight-binding approach is the method which is able to take
into account both the bulk and the interface contribution to the
spin-orbit interaction of 2D electrons by a rigorous uniform way
\cite{kp}. Tight-binding calculations of the spin splittings of 2D
electron in the [110] quantum wells  grown from III-V semiconductors
were recently performed in Refs.~\cite{Nestoklon12,Nestoklon16_2}.

In this paper we perform the tight-binding calculations of the spin
splittings of the electron energy spectrum in the [100] GaAs quantum
wells subjected in an electric field along the growth direction. We
compare the obtained dependencies of  the spin splittings on the
quantum well width and the electric field with the analytic
expressions derived within the one-band electron Hamiltonian
containing the bulk \cite{Pikus} and the two interface
\cite{Alekseev1,Alekseev2} spin-orbit terms. From the comparison, we
extract the values of the bulk and the interface parameters
in the effective electron Hamiltonian. The analytical one-band
calculations perfectly reproduce the results of tight-binding
numerical calculations for different quantum wells. As a result, we
estimate the interface terms and
prove the importance of the interface contributions to the spin
splitting of 2D electron spectrum in GaAs quantum wells.

\section{Tight-binding calculations}
Electron states in the quantum well structure are calculated in the
extended basis $sp^3d^5s^*$ tight-binding approach which is known as
an efficient empirical-parameter full-band representation of
semiconductor electronic properties \cite{Jancu98}. The coordinate
system is chosen in such a way that the cation atom is located at
the origin and one of its neighbors is located in [111] direction.
This choice results in the opposite sign of cubic
spin splitting constant \(\gamma_c\) and the constants of linear
spin splitting in comparison with the 
``anion in the origin'' convention.

We consider GaAs quantum well between Ga$_{0.7}$Al$_{0.3}$As
barriers. The alloys are treated in the virtual crystal
approximation: all tight-binding parameters are taken as weighted
linear combination of the corresponding GaAs and AlAs parameters,
which means that we neglects possible effects of bowing and
disorder. This is a good approximation for GaAs/AlGaAs
heterostructures, but for other alloys a more sophisticated
approximation might be necessary \cite{Nestoklon16}.

The tight-binding parameters are taken from
Ref.~\onlinecite{Jancu98}. To calculate the spin splitting we choose
small wave vector $\bm{k}$ and change its direction in the (001)
plane. At finite $\bm{k}$, the (double) degeneracy of
quantum-confined electron states is lifted, with the splitting
proportional to $|\bm{k}|$. For each lateral direction of $\bm{k}$,
we calculate the splitting $\Delta(\bm{k})$ and the vector $\bm{s}$
of the mean value of electron spin for the lower spin branch. The
typical splitting is of the order of meV. We note that the
tight-binding method provides high accuracy of the spin splitting
near the band edge \cite{Jancu05,Nestoklon06,Nestoklon08} since
$\Delta(\bm{k})$ is determined by the difference between the
energies of spin subbands and possible inaccuracy in the band
positions does not affect its value significantly. The electric
field $E_z$, applied along the growth direction and causing
the quantum well asymmetry, is taken into account in the
framework of standard procedure \cite{Graf95} by shifting of
the diagonal energies due to the local potential at atomic
sites.

\begin{figure}
\epsfxsize=240pt {\epsffile{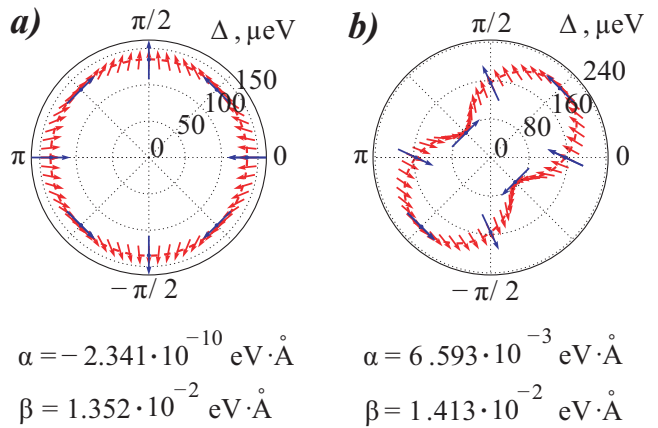}}
 \caption{ The 2D electron spin splittings and
the  direction of the mean spin values as  functions of
lateral wave vector angle for $|k|=5\cdot 10^{-3}$\AA$^{-1}$
calculated for quantum well width 30AL. The results for
tight-binding calculations (long blue arrows) are compared with
results obtained in one-band Hamiltonian \eqref{eq:H_eff} (dashed
curve for spin splitting and short red arrows for spin direction).
Left panel shows results for zero electric field and right panel for
the electric field $1\cdot 10^5$eV/cm.
 }\label{fig:TB}
\end{figure}

The effective Hamiltonian for an electron in quantum well can
be written as:
\begin{equation}
\label{eq:H_eff}
 \mathcal{H}_{\text{eff}}(\bm{k}) = \beta \left( k_x \hat{\sigma}_x  - k_y \hat{\sigma}_y  \right) +
   \alpha\left( k_y \hat{\sigma}_x  - k_x \hat{\sigma}_y  \right)
   \:.
\end{equation}
The solution of this one-band Hailtonian gives us the splitting
between the two states:
\begin{equation}
\label{eq:Delta_eff}
  \Delta(\bm{k}) = 2 \sqrt{(\beta^2 + \alpha^2)k^2 + 4 \alpha\beta k_x k_y}
  \:,
\end{equation}
 and the mean value of spin projection for the lower energy state:
\begin{equation}
\label{eq:s_eff}
\begin{split}
  s_x^{\text{eff}} &= - \frac{\beta k_x + \alpha k_y}{\Delta(\bm{k})}
  \:,
  \\
  s_y^{\text{eff}} &= \frac{\beta k_y + \alpha k_x}{\Delta(\bm{k})}
  \:.
\end{split}
\end{equation}
The splittings and spin direction as functions of the lateral wave
vector angle are extracted from tight-binding calculations are
fitted with equations \eqref{eq:Delta_eff}, \eqref{eq:s_eff}. The
fit is shown in Fig.~\ref{fig:TB}. This allows us to extract the
Dresselhaus and Rashba constants $\beta$ and $\alpha$ directly from
the tight-binding calculations.

In accordance with symmetry consideration, the parameter $\alpha$
vanish at zero electric field, when the quantum well is symmetric,
and then increase linearly with $E_z$. The Dresselhaus parameter
$\beta$ depends on the electric field $E_z$ in a more weak
manner: it starts to deviate significantly from zero-field value
only when the variation of electrostatic potential $-eE_z z$ from the
electric field $E_z$ in  the interface regions  is comparable with
the quantum confinement energy.

Repeating the calculation procedure for different quantum wells,
we obtain the dependence of the spin-orbit coupling
parameters on the quantum well width. The dependence of the
Dresselhaus parameter $\beta$ on the quantum well width is
non-monotonic. This is expected for $\bm{k}$-linear splitting caused
only by $\bm{k}$-cubic terms in the bulk crystal. We discuss
this point in more details below  within the one-band electron
Hamiltonian containing the bulk as well as the interface spin-orbit
terms.

\section{Spin splittings within  the one band approach}
The cubic term in the Hamiltonian of zincblende semiconductor
\cite{Pikus} in a quantum well grown along [001] direction may be
written in linear in the lateral wavevector order as:
\begin{equation}
\label{Dr}
 \hat{H}_{BIA}=\frac{d}{dz} \gamma(z)\frac{d}{dz}
(k_x\hat{\sigma}_x - k_ y \hat{\sigma}_y) \:,
\end{equation}
Here $\gamma(z)$ is the bulk spin-orbit parameter, which depends on
the layer material. Analysis of the experimental data for bulk GaAs
yield the value of  $\gamma$  around $25$~eV$\cdot${\AA}$^3$
\cite{Pikus}.

The interface contributions to the spin-orbit interaction of  2D
electrons in a quantum well have the same symmetry relative to the
symmetry operations in the $xy$ plane, but instead the operator
$d^2/dz^2$ they contains  the delta-function and its derivative
localized at the well interfaces
\cite{Rossler,Alekseev1,Alekseev2,13}:
\begin{equation}
\label{int}
\begin{array}{c}
\hat{H}_{int}=\sum \limits _{\nu=l,r } (\hat{H}_{int,0, \nu}+ \hat{H}_{int,1,\nu})
\:,
\\
\\
\hat{H}_{int,0,\nu} = \zeta_{\nu}
\delta(z-z_{\nu}) (k_x\hat{\sigma}_x - k_ y \hat{\sigma}_y)
\:,
\\
\\
\hat{H}_{int,1,\nu} = \xi_{\nu}
\delta'(z-z_{\nu}) (k_x\hat{\sigma}_x - k_ y \hat{\sigma}_y)
\:.
\end{array}
\end{equation}
The parameters  $\zeta_{l,r}$ and  $\xi_{l,r}$ are determined by the
structure of the chemical bonds of the atom at the interface.
Comparison of the results of tight-binging calculation and the
analytical expressions  (\ref{int}) allows us to determine $\zeta$
and  $\xi$.

The equations (\ref{int}) assumes continuity of the wave function
derivative at the interfaces of the quantum well. If we use the
one-band electron Hamiltonian with different values of the effective
mass in the barrier and in the well, then the wave function
derivative at the quantum well interfaces is discontinuous, and we
should modify the form of the interface  contribution (\ref{int}).

The equations (\ref{int}) also assumes some fixed form of the bulk
spin-orbit term containing discontinuities of the bulk Dresselhaus
parameter $\gamma(z)$ at the well interfaces. 
Other forms of the Hamiltonian (\ref{Dr})  are
allowed. For example, one can take them proportional to
$[\gamma(z) \,d^2/dz^2 + d^2/dz^2 \gamma(z) ]/2$ or
$\gamma^{1/2}(z) d^2/dz^2 \gamma^{1/2}(z) $. Our choice is based
only on the reason of most simplicity of  Eq. (\ref{Dr}). Of course,
this freedom of choice is limited by the requirement  that Eq.
(\ref{Dr}) should be an Hermitian operator.

Let $u(z)$ be the electron wavefunction of the first level of space
quantization within the one-band electron Hamiltonian. Projection of
the operators (\ref{Dr}) and (\ref{int}) onto the first subband,
 corresponding to the  wave function $u(z)$, leads to the following form of
the bulk and the surface contributions to the anisotropic spin-orbit
interaction of 2D electrons:
\begin{equation}
\label{2D}
\begin{array}{c}
 \displaystyle
\hat{H}_{D}=\beta \,  (k_x\hat{\sigma}_x - k_ y \hat{\sigma}_y)
\:,
\\
\\
 \displaystyle
\beta  = -  \int \limits_{-\infty}^{\infty} dz \: \gamma (z)[u'(z)]^2
+
\\
\\
 \displaystyle
+ \sum \limits _{\nu=l,r} \left\{ \zeta_{\nu} u(z_{\nu})^2 +
\xi_{\nu} [u(z_{\nu})^2]' \right\}
 \:.
\end{array}
\end{equation}

We consider a rectangular quantum well.
The parameter $\gamma(z)$ in
this structure is a step-like function with the two different
values in the well and in the barrier layers:
\[
\gamma(z)=
 \left|
 \begin{array}{l}
\gamma_b \:, \;\; z<0 \, ,\; z>a
 \\
 \gamma_w \:, \;\; 0<z<a
 \end{array}
 \right.
 \:.
\]
Without an electric field the symmetry of the quantum well $D_{2d}$
leads to relations between the coefficients $\zeta_{\nu}$ and
$\xi_{\nu}$ for the left and right interfaces:
$\zeta_{l}=\zeta_{l}$, $\xi_{l}=-\xi_{r}$. The wave function $u(z)$
is symmetric  with respect to reflection plane in the center of the
well, and we obtain
\begin{equation}
\label{symm_well}
\begin{array}{c}
 \displaystyle
\beta  = -  \int \limits_{-\infty}^{\infty} dz \:  \gamma (z)
[u'(z)]^2 + 2 \zeta u(z_{r})^2 +  2  \xi [u(z_{r})^2]' \:.
\end{array}
\end{equation}

If an electric field $E_z$ is applied along the normal to the
quantum well, the relationships $\zeta_{l}=\zeta_{l}$,
$\xi_{l}=-\xi_{r}$ are  retained, but the wave function $u(z)$ is
not symmetric with respect to the quantum well center. In this case,
\begin{equation}
\label{asymm_well}
\begin{array}{c}
 \displaystyle
 \beta  = \beta_{b}+ \beta_{int,0} +\beta_{int,1}
 \:,
 \\
 \\
 \displaystyle
\beta_{b} =  -  \int \limits_{-\infty}^{\infty} dz \:
 \gamma (z) [u'(z)]^2
 \:,
\\
\\
 \displaystyle
 \beta_{int,0} =
  \zeta \, [u(z_{l})^2+u(z_{r})^2]
  \:,
  \\
  \\
  \displaystyle
  \beta_{int,1} = \xi \,
\{[u(z_{l})^2]'-[u(z_{r})^2]'\}
 \:.
\end{array}
\end{equation}

The potential energy of an electron in an empty  rectangular quantum
well in a homogenous electric field $E_z$ is:
\begin{equation}
\label{true}
 U(z) = eE_zz + \left|
\begin{array}{l}
0, \; \;  0  < z  <  a
\\
U_0, \; \;  z < 0 , \; z  >  a
\end{array}
\right.
\end{equation}
Here $e>0$ is the absolute value of the electron charge. In order to
calculate the wave function $u(z)$ analytically we substitute the
real potential (\ref{true}) of the  quantum well by  the model
potential
\begin{equation}
 \label{model}
 \widetilde{U}(z) =
\left|
\begin{array}{l}
eE_zz   \: , \; \; \;  0 < z  <  a
\\
U_0 \: , \;\;  \; z  <0
\\
U_0 +  eE_za \: , \;\;\;  z    > a
\end{array}
\right.
\: .
\end{equation}
The electron ground level $E_0$ is resonant  for the true potential
$U(z)$ (\ref{true}), but turns into actually stationary level for
the model potential $\widetilde{U}(z)$ (\ref{model}). We assume that
the well is deep: $E_0 \ll  U_0 $.  The approximation of
Eq.~(\ref{model}) is valid in the electric fields $E_z$, for which
the condition $eE_z /\kappa \ll U_0$ is fulfilled, where $\kappa =
\sqrt{2m(U_0-E_0)}/\hbar$ is the reciprocal length of decay of the
electron wave function in the barrier.

The wavefunction $u(z)$ of the ground state  corresponding to the
model potential $\tilde{U}(z)$ is expressed via the Airy functions
in the well  region:
\begin{equation}
 \label{psi_well}
 u(z)=c_A \mathrm{Ai}\left( \frac{z-E_0/e E_z}{d}
\right)
 + c_B \mathrm{Bi}\left( \frac{z-E_0/e E_z}{d} \right)
\end{equation}
and the exponents  in the barriers:
\begin{equation}
 \label{psi_barr}
\begin{array}{l}
u(z)=c_l e^{\kappa_l z},\;\; z<0 \:,
\\
\\
u(z)=c_r e^{-\kappa_r \,(z-a)},\;\; z>a \:.
\end{array}
\end{equation}
Here $d=(2meE_z/\hbar^2)^{-1/3}$,  $\kappa_l = \kappa
=\sqrt{2m(U_0-E_0)}/\hbar$,
$\kappa_r=\sqrt{2m(U_0+eE_za-E_0)}/\hbar$. The coefficients $c_A$,
$c_B$, $c_l$, $c_r$, and the eigenenergy $E_0$ were calculates as
the functions of the quantum well width and the electric field $E_z$
by use of the standard methods.

\begin{figure}
\epsfxsize=240pt {\epsffile{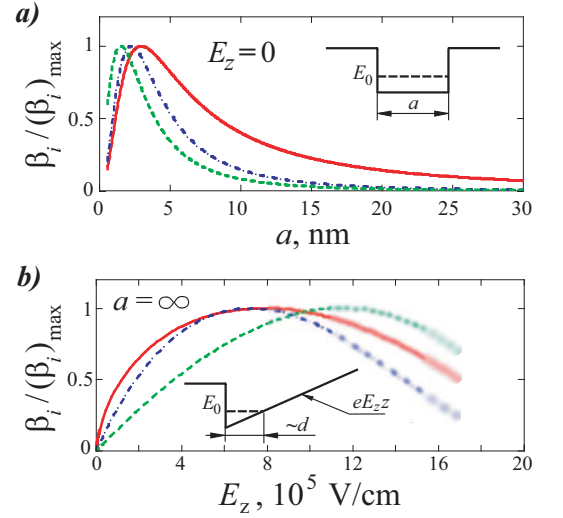}}
 \caption{Magnitudes $\beta_i$,  $i=b, \, (int,0),\, (int,1) $,
of the different contributions  to the anisotropic part of the 2D
electron spin-orbit interaction  as functions of the quantum well
width for zero electric field (a) and  as functions of electric
field  for an infinitely large  quantum well width (b). Red solid
lines correspond to the bulk Dresselhaus (BIA) contribution
$\beta_b$, while  blue dash-dot and green  dash lines correspond to
the interface (IIA) contribution $\beta_{int,0}$ and
$\beta_{int,1}$, respectively. Insets on the both panels (a) and (b)
schematically demonstrate the  potential energy  of an electron in
the  quantum wells. All the dependencies $\beta_i(a)$  and
$\beta_i(E_z)$ are normalized to their maximum values within the
chosen intervals of the arguments $a$ and $E_z$.
}\label{fig:2}
\end{figure}

In Fig.~\ref{fig:2} we present  the bulk and the interface contributions to
the 2D electron spin-orbit interaction calculated by
Eqs.~(\ref{asymm_well}), (\ref{psi_well}), and (\ref{psi_barr}). In
this paper we consider  quantum wells in the typical
heterostructures
Ga$_{0.7}$Al$_{0.3}$As$/$GaAs$/$Ga$_{0.7}$Al$_{0.3}$, for which we
take the following electron band parameters: $U_0=$300~meV and
$m=0.067m_0$. The panel (a) of Fig.~\ref{fig:2} presents the dependencies of
the bulk and the interface contributions to the spin-orbit parameter
$\beta$ on the quantum well width in the absence of electric field
$E_z$ along the $z$ direction. We see that all the contribution have
maximums at the well widths $a\sim 2-3$~nm and tend to zero at very
small and very large $a$. The endings of the curves in the region of
small widths $a$ correspond to the one-monolayer quantum well
 for which  $a \sim  a_0$ and  the effective mass method and the spin-orbit
Hamiltonians (\ref{Dr}) and (\ref{int}) are not applicable
surely (here $a_0$ is the lattice constant). The dependencies
$\beta_{b} (a) $ and $\beta_{int,1,2}(a) $   exhibit the power
behavior in the region of large widths $a$:
\begin{equation}
 \label{ass_large_a}
 \beta_b \sim a^{-2}
 \:, \;\;
 \beta_{int,0} \sim a^{-3}
 \:, \;\;
  \beta_{int,1} \sim a^{-3}
   \:,
\end{equation}
as well as in the region of small widths $a$:
\begin{equation}
\label{ass_small_a}
 \beta_b \sim a^{2}
 \:, \;\;
 \beta_{int,0} \sim a
 \:, \;\;
  \beta_{int,1} \sim a^{2}
   \:.
\end{equation}
However   we must keep in mind that Eq. (\ref{ass_small_a}) imply
that the quantum well width is large enough, as minimum  $a \gtrsim
a_0$.

The panel (b) in Fig.~\ref{fig:2} shows the dependencies of the bulk and the
interface contributions to the spin-orbit parameter $\beta$  on the
magnitude of the electric field $E_z$ for the infinitely large
quantum well width. The absolute values of $\beta_{b} $ and
$\beta_{int,0,1} $  increase with the electric field $E_z$ in the
region from $E_z=0$ up to $E_z \sim U_0 \kappa / e$, where the
approximate electron potential (\ref{model}) is yet applicable. At
small electric fields the values $\beta_{b} $ and $\beta_{int,0,1} $
depends on electric field $E_z$ as:
\begin{equation}
\label{ass_Ez}
 \beta_b \sim (E_z)^{2/3}
 \:, \;\;
 \beta_{int,0} \sim E_z
 \:, \;\;
  \beta_{int,1} \sim E_z
  \:.
\end{equation}
At larger electric fields, $E_z > U_0 \kappa / e$, the energy level
$E_0$ becomes resonant to a large extent. This is illustrated in
Fig.~\ref{fig:TB}(b) by blur and broadening of the graphic lines.

\begin{figure}
\epsfxsize=240pt {\epsffile{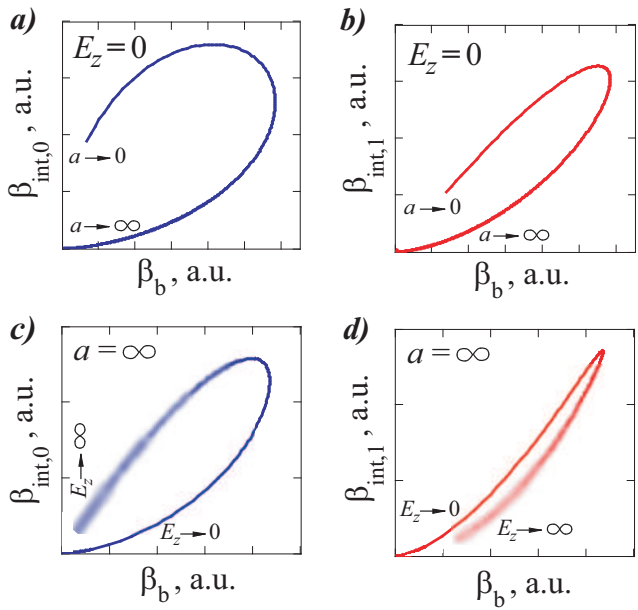}}
 \caption{ The parametric plots
of the bulk $\beta_{b}$ (BIA) and interface  $\beta_{int,0,1}$ (IIA)
contributions to the anisotropic part of the 2D electron spin-orbit
interaction.  Panels (a,b) demonstrate the positions of the points
$[\beta_{b},\beta_{int,(0/1)}]$ for the continuously varying quantum
well width $a$ at zero electric field  $E_z=0$, while  the panels
(c,d) show the positions of the points
$[\beta_{b},\beta_{int,(0/1)}]$ for the continuously varying
electric field $E_z$ and infinitely large quantum well width, $a \to
\infty$.
 }\label{fig:bilk_vs_int}
\end{figure}

To compare the dependencies of  the bulk and the interface
contributions  on the quantum well width and the electric field we
have drawn the parametric plots of the values $\beta_{int,0}$,
$\beta_{int,1}$, and $\beta_{b}$.  In the panels (a,b) of
Fig.~\ref{fig:bilk_vs_int} we plotted the pairs $[ \beta_{b} , \beta_{int,0}   ]$ and $
[\beta_{b} , \beta_{int,1} ]$ for continuously varying values of the
the quantum well width $a$ at zero electric field $E_z=0$. We see
that all contributions $\beta_{int,0}(a)$, $\beta_{int,1}(a)$, and
$\beta_{b}(a)$ are approximately proportional each other in 
the wide intervals of the values of quantum well width $a$. This
implies that if we have in hand the experimental or the numeric
(e.g., tight-binding) dependence $\beta(a)$ of the total spin
splitting, which has some uncertainty $\delta \beta(a)$, we can
establish the relative magnitudes of the bulk $\beta_b(a)$ and the
interface $\beta_{int,0,1}(a)$ contributions in this dependence
$\beta(a)$ only for rather small values of the uncertainty, $\delta
\beta(a) \ll \beta(a) $. Otherwise, if the uncertainty $\delta
\beta(a)$ is large, one cannot designate the contributions
$\beta_{b}(a)$, $\beta_{int,0}(a)$, and $\beta_{int,1}(a)$ in it as
the set of the functions $\beta_{b}$, $\beta_{int,0}$, and
$\beta_{int,1}$ is close to be  {\em a linearly dependent set}.
For  example, for large $a$, $a \to \infty $ one obtains from
Eq.~(\ref{ass_large_a}) that $\beta_{b} \to 0$,  $\beta_{int,0,1}
\to 0$,   and $\beta_{int,0,1} \sim \beta_{b}^{3/2} $, see
Fig.~\ref{fig:2}(a,b).

In the panels (c) and (d)  of Fig.~\ref{fig:bilk_vs_int} we show the parametric
plots of the pairs $ [\beta_{b} , \beta_{int,0} ]$ and $[\beta_{b} ,
\beta_{int ,1} ]$ for the infinitely wide quantum well and
continuously varying values of the electric field $E_z$ from zero up
to the limit value $\sim U_0\kappa/e$. We again see that the
functions $\beta_{b}(E_z)$, $\beta_{s,0}(E_z)$, and
$\beta_{s,1}(E_z)$ are almost proportional.
At small $E_z$, $E_z \to 0$, one obtains from Eq.~(\ref{ass_Ez})
that $\beta_{b} \to 0$, $\beta_{int,0,1} \to 0$, and
$\beta_{int,0,1} \sim \beta_{b}^{3/2} $, see Fig.~\ref{fig:2}(c,d).

It should be mentioned also that for a quantum well with a
substantial concentration of electrons, the electron potential
energy  $U(z)$ is strongly modified by the presence of
electron charge inside quantum well. The potential  $U(z)$ should be
calculated simultaneously with the energy levels in  a
self-consistent procedure of a joint solution of the Poisson and
Schr\"odinger equations. At the edges of the quantum well  the
resulting wavefunction has the values, which are  significantly
greater by their absolute values than for the wave function
in the quantum well   with the same width and depths, but containing
no electrons (for example, see Ref.~\cite{Alekseev2}). This means
that the large electron concentration in a quantum well  leads to an
increase of  the role of the interface contributions to the
spin-orbit interaction compared with an empty quantum well.

\section{Analysis of relative importance of different contributions  }
\begin{figure}
\epsfxsize=220pt {\epsffile{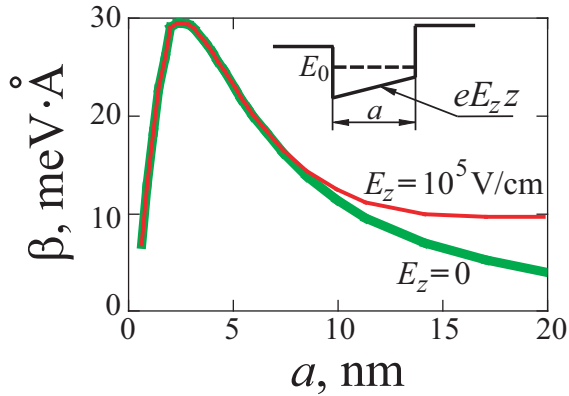}}
 \caption{ The results of tight-binding calculations of the 2D
electron spin splitting parameter $\beta$ as a function of the
quantum well width $a$ for the electric fields $E_z=0$  and
$E_z=10^5$~V/cm .
 }\label{fig:4}
\end{figure}

With the help of the tight-binding approach, we have
calculated the absolute values of the spin splitting constants as
functions of the well width in the range from 1 to
20~nm, and the electric fields $E_z$ in the range
from from 0 up to $10^5$~V$/$cm. The results of these calculation
for $E_z=0$ and $E_z=10^5$~V$/$cm are shown in Fig.~\ref{fig:4}.

The character electric field $E_{z,\mathrm{max}}$ in which  the
center of the wavefunction $u(z)$ becomes substantially shifted from
the quantum well center should be calculated from the equality $
eE_za \sim E_0|_{E_z=0}$.  For the quantum well with the parameters
$a=10$~nm and $U_0=300 $~meV we obtain $E_{z,\mathrm{max}}\sim 3
\cdot 10^4$~V$/$cm.

\begin{figure}
\epsfxsize=220pt {\epsffile{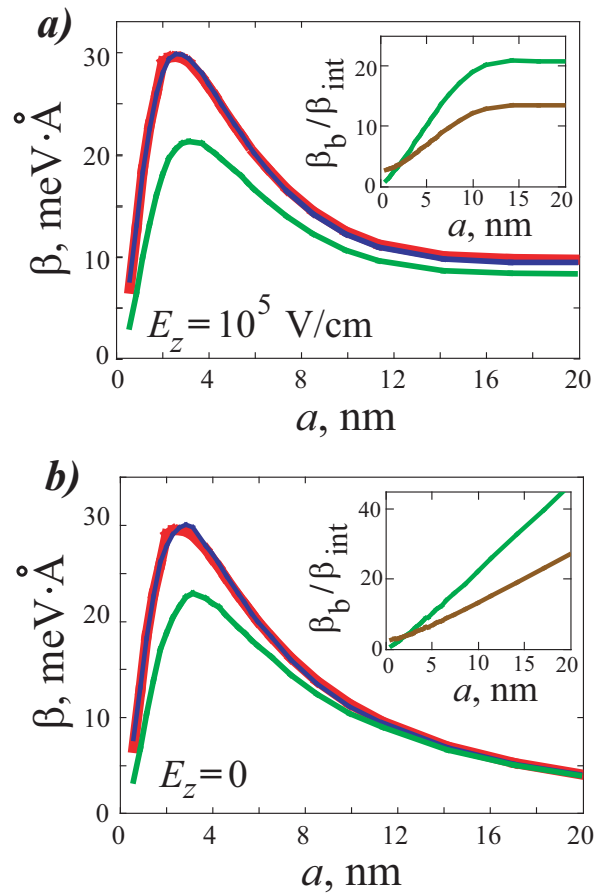}}
 \caption{The 2D electron spin
splitting parameter $\beta$ as a function of the quantum  well
width. Panel (a) presents the results for the finite electric field
$E_z=10^5$~V/cm, while the panel (b) corresponds to $E_z=0$~V/cm.
Red lines are the results of the  tight-binding  numerical
calculations. Blue curves represent the analytical one-band
calculations, Eqs.~(\ref{symm_well},\ref{asymm_well}).
The parameters $\gamma$, $\xi$, and $\zeta$ are
fixed for all values of electric field; they correspond to
minimal deviation from the tight-binding result for the electric
field $E_z=10^5$~V/cm (the red curve at panel (a)). The green curves
are the bulk Dresselhaus contributions $\beta_b (a)$  (with the
parameter $\gamma$ obtained  in the minimizing procedure).
Insets exhibit the ratios $\beta_b/\beta_{int,0}$
and $\beta_b/\beta_{int,1}$ of the interface and the bulk
contributions to the spin splitting parameter $\beta$ as functions
of the well width $a$.}\label{fig:5}
\end{figure}

We fitted the obtained dependence $\beta(a)$  for $E_z=10^5$~V$/$cm
by the analytical dependence (\ref{asymm_well}) by the least square
method. We take into account that the bulk spin-orbit Dresselhaus
parameter $\gamma$ is different in the well and in the barrier
regions of the heteostructure. As we consider the heterostucture
with the barriers grown  from the compound Al$_{0.3}$Ga$_{0.7}$As,
we adopted the following estimation of the ratio  of the spin-orbit
bulk parameters in the well and in the barrier:
$\gamma_{bar}/\gamma_{w}=0.7$. It is seen from the Fig.~\ref{fig:5}(a) that the
numerical dependence $\beta(a)$ can be very well reproduced by the
analytical formula (\ref{asymm_well}) with the three parameters
$\gamma$, $\xi$, $\zeta$. From the fitting procedure we obtained the
following values of the bulk and interface parameters:
$\gamma_{GaAs}=-23$~eV$\cdot ${\AA}$ ^3$, $\xi=-1.5$~eV$\cdot
${\AA}$ ^3$, $\zeta =6.5 \cdot 10^{-6}$~eV$\cdot ${\AA}$ ^2$.

We checked that, for the obtained parameters $\gamma$, $\xi$,
$\zeta$, the analytical one-band dependencies  $\beta(a)$ for all
the electric fields $E_z$ in the interval from 0 up to $10^5$~V$/$cm
well coincides with the results of the tight-binding calculations of
$\beta(a)$. In the Fig.~\ref{fig:5}(b) we show the  numeric tight-binding
and the analytical  one-band functions $\beta(a)$ for $E_z=0$. These
two curves almost coincide. The obtained  result proves that the
description of the electron spin-orbit interaction in the GaAs
quantum well with sharp interfaces within the one-band Hamiltonians
(\ref{Dr}) and (\ref{int}) is adequate.

The obtained  value of the coefficient  $\xi$ in the interface term
$\hat{H}_{int,1,\nu}$ 
for GaAs quantum wells
corresponds by the order of magnitude to the
value of the interface contribution in linear spin splitting 
which has been estimated within  the framework of 
multi-band {\bf k}$\cdot${\bf p}-Hamiltonianin in 
Refs.~\cite{Rossler,Glazov_priv}.

\section{Conclusion}
In conclusion, we show that the results of atomistic calculations of
spin splittings of the 2D electron spectrum in GaAs quantum wells
can be perfectly reproduced in the framework of one-band
effective-mass model by adding the interface terms in the
one-band Hamiltonian.  By introducing the two independent
parameters in  the interface terms, the microscopic atomistic
calculations are reproduced with the correct functional dependencies
of spin-dependent terms on the quantum well width and the
electric field applied.

The effective one-band description allows us to conclude that the
interface-induced anisotropy contributes significantly to the value
of the coefficient in the Dresselhaus term in the electron
Hamiltonian in quantum wells.

We have also demonstrated that the separation of bulk and interface
terms in the experiments is complicated by the fact that both terms
contribute to the linear spin splitting of the same symmetry and
have very similar functional dependency on quantum well width and
electric field applied.

It should be noted that, in the real quantum wells, there also
exists  a contribution to the observed magnitude of the spin-orbit
interaction from the electron-electron interaction
\cite{Krishtopenko}. The strength of this interaction inducing
renormalization of the constant $\beta$ depends on the geometry of
the quantum well and the electron density. This fact more
complicates the interpretation of the experimental data on the
absolute value of the spin-orbit coupling in quantum wells.

\section*{Acknowledgments} 
This work was supported by RFBR grants 14-02-00168 and 15-02-06344, 
by the Russian Ministry of Education and Science 
(Contract No. 14.Z50.31.0021, Leading scientist: M. Bayer), 
and by the Dynasty Foundation.

\end{document}